\documentclass[doublecol]{epl2} 

\title{Probing Majorana and Andreev Bound States with Waiting Times}
\shorttitle{Probing Majorana and Andreev Bound States with Waiting Times} 

\author{D. Chevallier\inst{1} \and M. Albert\inst{2} \and P. Devillard\inst{3}}
\shortauthor{D. Chevallier \etal}

\institute{                    
  \inst{1} Department of Physics, University of Basel, Klingelbergstrasse 82, CH-4056 Basel, Switzerland\\
  \inst{2} Universit\'e C\^ote d'Azur, CNRS, INLN, France\\
  \inst{3} Aix Marseille Univ, Univ Toulon, CNRS, CPT, Marseille, France
}

\abstract{
We consider a biased Normal-Superconducting junction with various types of superconductivity. Depending on the class of superconductivity, a Majorana bound state may appear at the interface. We show that this has important consequences on the statistical distribution of time delays between detection of consecutive electrons flowing out of such an interface, namely the waiting time distribution. Therefore, this quantity is shown to be a clear fingerprint of Majorana bound state physics and may be considered as an experimental signature of its presence.
}

\begin{document}

\maketitle

\section{Introduction} 
\begin{figure}
  \includegraphics[width=0.95\linewidth]{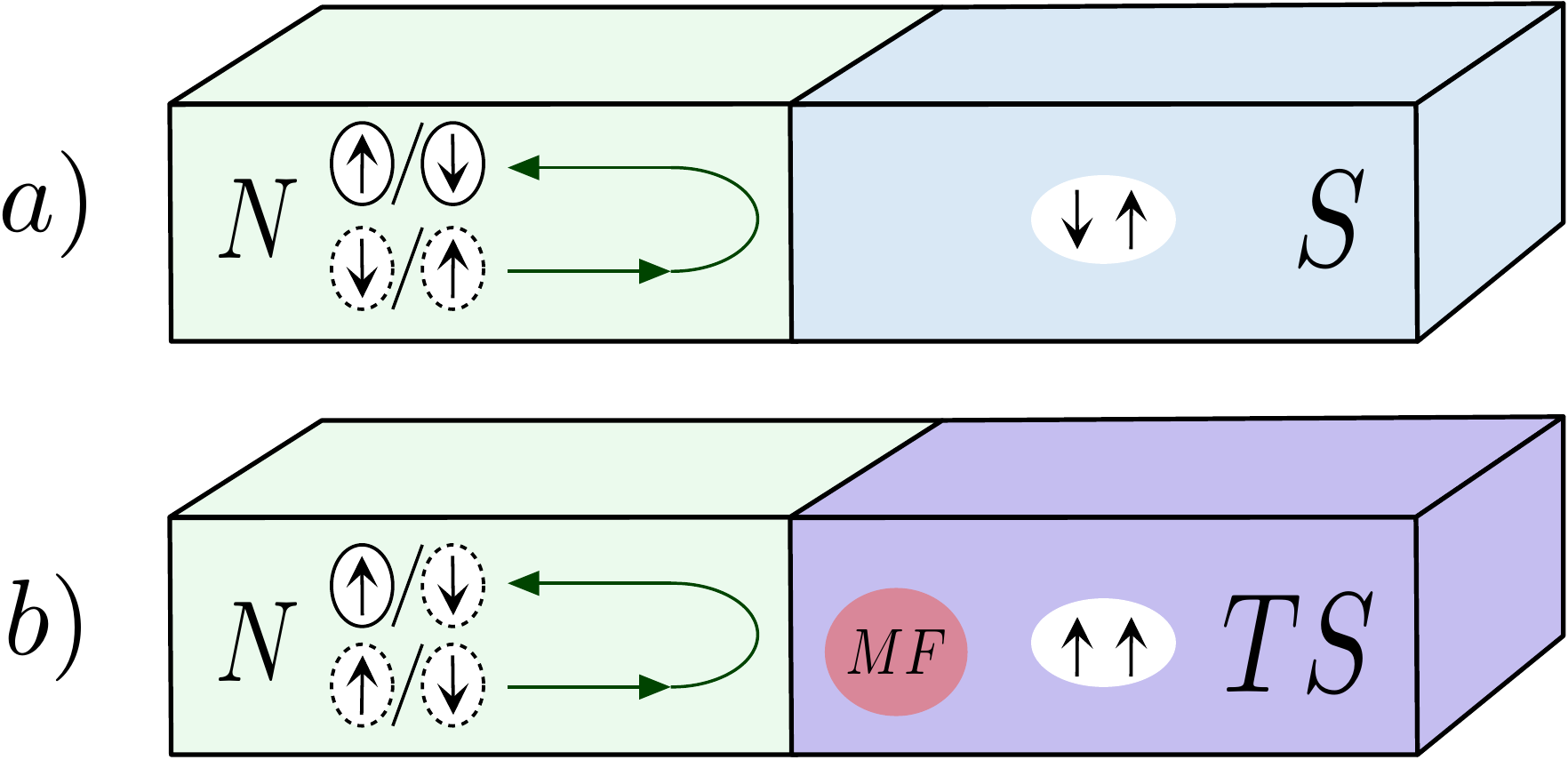}
  \caption{(color online) Schematic picture of the Andreev reflection processes in the two different junctions. a) Normal-(trivial)Superconducting junction: a hole with spin $\uparrow_{\hat{n}}$ ($\downarrow_{\hat{n}}$) is converted into an electron with opposite spin $\downarrow_{\hat{n}}$ ($\uparrow_{\hat{n}}$) and b) Normal-(topological)Superconducting junction: a hole with spin $\uparrow_{\hat{n}}$ is converted into an electron with same spin $\uparrow_{\hat{n}}$  and a hole with spin $\downarrow_{\hat{n}}$ is reflected as a hole also with the same spin $\downarrow_{\hat{n}}$. In the first case $\hat n$ denotes any possible direction whereas in the second case a Majorana bound state appears at the interface and sets a special spin direction $\hat n$ for scattering (see text). In both cases a bias voltage $eV$ is imposed and brings the superconducting chemical potential $\mu_S$ above the Fermi energy $E_F$ of the normal metal with the restriction that $eV\ll\Delta$ the superconducting gap.\label{fig_setup}}
  \label{fig:Fig1}
\end{figure}

During the last two decades, Majorana fermionic states in condensed matter physics, have received a lot of interest because of their exotic properties such as non-Abelian statistics, that open the perspective of using them for quantum computation. These exotic states have been studied extensively in various systems \cite{Fu,MF_Sato,MF_Sarma,MF_Oreg,alicea_majoranas_2010,potter_majoranas_2011,Klinovaja_CNT,Pascal,Bena_MF,Fra,Rotating_field,Ali,RKKY_Basel,RKKY_Simon,RKKY_Franz,Pientka,Ojanen,Carlos} with among them, a conceptually simple one made with a semiconducting nanowire of InAs or InSb, with strong spin-orbit coupling, subjected to an external Zeeman field and in the proximity of an s-wave superconductor (SC) \cite{MF_Sarma,MF_Oreg,Alicea,Beenakker}. In this situation, a Majorana Bound State (MBS) may appear at the interface of a normal/superconducting junction, under proper conditions, and strongly affects the electronic conduction properties (see Fig. \ref{fig:Fig1}). Several experiments have reported the observation of a zero-bias conductance peak in such physical setups, which are in good qualitative agreement with all theoretical predictions based on Majorana physics so far but still not fully consistent with the predicted conductance and magnetic field value needed for the existence of a MBS \cite{Mourik,Marcus,Heiblum,Us}. Therefore, several works have been conducted in order to understand these inconsistencies based on alternative interpretations by including other physical processes \cite{DeFranceschi,Liu2012,Atland2012,Pientka2012,Pikulin2012}. However, a clear consensus is still lacking mostly because of the absence of an experimental smoking gun for Majoranas. Generally these MBS appear in hybrid junction by tuning one of the parameter of the system (i.e. phase difference or Zeeman field) in a topological phase. Along this transition, these states mutate from Andreev Bound State (ABS) in the non-topological phase to MBS in the topological one and understanding their differences is thus of fundamental importance in order to distinguish them. So far, many efforts have focused on the relation and the evolution of ABS onto MBS by tuning the system parameters \cite{Pascal,Klinovaja,Ramon} but less on their own properties \cite{Beenakker,Fu2009,Rubbert,Beenakker2016} and the consequences on physical observables which is the purpose of this contribution.

Recently, an intriguing feature due to MBS was identified in Ref. \cite{He2014} and named selective equal-spin Andreev reflection (SESAR). The presence of a MBS drastically modifies Andreev reflection and leads to a spin polarization of the current as well as to interesting correlations between different spin components which are visible in the zero frequency noise \cite{Haim2015} for instance. However, such fingerprints are based on the possibility to observe fine quantitative differences between spin resolved current-current cross correlations which seems to be complicated experimentally in the present situation. In this letter, we show that a very clear qualitative difference is visible in the Waiting Time Distribution (WTD) of electrons flowing out of the interface making it an interesting and alternative signature of MBS. The WTD is the statistical distribution of time delay between the detection of two consecutive electrons and has been shown to be a very informative and powerful quantity for understanding correlations in mesoscopic quantum conductors \cite{Brandes2008,Albert11,Albert12,FlindtThomas,Dasenbrook,Rajabi2013,FlindtThomas2,Sothmann,Albert2014,Haack2014,Hofer16,Albert16,Dasenbrook2016,Brandes2016}.

\section{Model} We consider two types of hybrid junctions as depicted on Fig. \ref{fig:Fig1} at zero temperature. The first one is a normal metal(N)/s-wave superconductor(S) carrying an ABS at the interface and the second one is a N/topological superconductor(TS) where the topological junction is made of a Rashba nanowire in proximity with an s-wave superconductor and in presence of Zeeman field $V_z$ carrying a MBS at each boundary. However, we assume the nanowire to be long enough to decouple the two MBS. In both situations, the Fermi energy of the normal metal is $E_F$, the superconducting gap is $\Delta$ and the superconducting chemical potential $\mu_S$ is biased in such a way that $E_F=\mu_S-eV$ and $eV\ll\Delta$ as shown on Fig. \ref{fig:Fig1}. As a consequence, a stream of non-interacting holes is approaching the interface from the Normal part where it is scattered as a coherent superposition of electron and holes. This incoming scattering state reads

\begin{equation}\label{eq:psiin}
  |\psi_{{\rm in}}\rangle=\prod_{k=0}^{k_{V}}c_{k,\hat{n}}c_{k,-\hat{n}}| 0\rangle,
\end{equation}
where $c_{k,\hat{n}}$ is the destruction operator of electron with momentum $k$, energy $E=\hbar v_F |k|$ and spin orientation $\hat n$ (or creation of holes with opposite properties), $| 0\rangle$ stands for the Fermi sea filled with states of energies up to $\mu_s$ and $k_V=eV/\hbar v_F$ with $v_F$ the Fermi velocity. So far, $\hat{n}$ denotes any unitary vector and not necessary $\hat z$ or $\hat x$ for instance. Below we will connect it to the polarization axes of the MBS but up to now this is just a choice of basis. It is important to note that spin components $\uparrow_{z}/\downarrow_{z}$ or more generally $\uparrow_{\hat{n}}/\downarrow_{\hat{n}}$ are equally distributed namely the incoming quantum state is isotropic in spin space. The key difference between these two junctions is how Andreev reflection occurs. In the case of a NS junction, the usual Andreev reflection takes place, meaning that a hole with a given spin is reflected as an electron with an opposite spin leading to the presence of ABS in such a junction (See Fig. \ref{fig:Fig1} a)). Replacing the s-wave superconductor by a topological one strongly changes scattering properties and especially the Andreev reflection. If the Zeeman field is strong enough to enter the topological phase ($V_z\ge \Delta$), the p-wave pairing dominates and the Andreev reflection is spin selective \cite{He2014} meaning that a hole with spin up is reflected as an electron with the same spin and a hole with spin down is normally reflected as a hole with spin down (See Fig. \ref{fig:Fig1} b)). More precisely, the presence of a MBS in the latter case, leads to a spin scattering symmetry breaking. There is a special spin orientation $\hat n$, called the Majorana's polarization, along which electrons or holes are totally Andreev reflected as a hole or electron respectively with spin conservation whereas particles with opposite spin are normally reflected. This is the essence of SESAR effect \cite{He2014} that leads to spin polarized current in this kind of hybrid junction. However, this precise direction cannot be determined from first principles and, in general, incoming particles are not spin oriented along this direction which leads to formally more complicated scattering although everything can be understood by decomposing the state onto this spin basis.

In order to evaluate the WTD and use it as a tool to probe the scattering properties of ABS and MBS in hybrid junctions we now need the expressions of the different outgoing scattering states. In such a junction, the interface plays an important role on the transmission which gives a finite width to the states \cite{PikulinJETP,PikulinPRB,Jose2016}. In Appendix A, we discuss this effect. However, for the sake of simplicity we focus on the zero temperature and perfect Andreev reflection limit and following Ref. \cite{Albert16,He2014} write down the different out-going quantum states.

\section{Outgoing states for N/S junction} In this case, the interface acts as a perfect Andreev mirror where all the holes with a given spin are Andreev reflected as electrons with opposite spin \cite{Albert16}

\begin{equation}\label{eq:psioutabs}
  |\psi_{{\rm out}}^{\textrm{\tiny{ABS}}}\rangle=\prod_{k=0}^{k_V}c^\dagger_{-k,-\hat{n}}c^\dagger_{-k,\hat{n}}| 0\rangle.
\end{equation}
Again, this quantum state is isotropic in spin space and simply corresponds to a stream of one-dimensional free electrons with energies between $\mu_s$ and $\mu_s+eV$ and two possible spin states (two channels of free fermions).

\section{Outgoing states for N/TS junction} The presence of the MBS deeply affects the outgoing state. As mentioned before, the key point is that a hole with a spin $\hat{n}$ is totally reflected as an electron with the same spin whereas a spin $-\hat{n}$ hole is subjected to perfect specular reflection with the same spin as well. It is therefore simpler to write the outgoing state in the Majorana polarization frame which reads

\begin{equation}\label{eq:psioutmbs}
|\psi_{{\rm out}}^{\textrm{\tiny{MBS}}}\rangle=\prod_{k=0}^{k_V}c^\dagger_{-k,\hat{n}}c_{-k,-\hat{n}}| 0\rangle,
\end{equation}
where it is now obvious that the outgoing stream of electrons is totally spin-polarized in the $\hat n$ direction. Therefore, if one were able to measure the electronic current in this spin direction, one would get the result of a perfect single quantum channel. On the contrary, if one measures it in a random direction, for instance $\hat z$, one would measure partition noise just because the spin operators $\hat S_z$ does not commute with $\hat S_{\hat n}$. In that sense, such an experiment is very similar to the so called Stern and Gerlach historical experiment as we will discuss later. Apart from this remark the many body state can be simply obtained from (\ref{eq:psioutmbs}), by replacing the $c^\dagger_{\hat n}$ with the proper linear combination of $c^\dagger_\uparrow$ and $c^\dagger_\downarrow$ of the right basis. This can be done using the scattering matrix obtained in \cite{He2014} depending on the value of the Zeeman field $V_z$. However, in the simple case where the Zeeman field is large compared to the superconducting gap, the Majorana is fully polarized in the z-direction which means that $\hat n$ and $\hat z$ are the same. Another simple case is right at the topological transition when the Zeeman field is just above the gap where $c^\dagger_{-k,\uparrow_{\hat n}}=\frac{1}{\sqrt{2}}(c^\dagger_{-k,\uparrow_{\hat z}}+c^\dagger_{-k,\downarrow_{\hat z}})$. Finally, it is important to note that in the non-topological case, namely when $V_z \ll \Delta$, the outgoing state behaves like in the s-wave junction (usual Andreev reflection on an ABS) \cite{He2014} and along the transition, the scattering matrix is not continuous.

\section{Waiting time distribution} We now turn to the calculation of the WTD. To do so, we need to specify the detection process. A time-resolved single electron detector is placed far away from the interface and is assumed to be sensitive to electrons only with energy above the superconducting chemical potential $\mu_S$. In addition the detector can be spin selective or not. Following Ref. \cite{Haack2014,Albert16} the WTD is obtained from the Idle Time Probability (ITP), namely the probability of not detecting any electron during a time slot $\tau$. The precise definition of it depends on the detector capabilities. Without spin filtering it reads

\begin{equation}\label{eq:Pi}
\Pi(\tau)=\langle \psi_{\rm out}| : e^{-Q_{\uparrow,E>\mu_s}} : : e^{-Q_{\downarrow,E>\mu_s}} : |\psi_{{\rm out}}\rangle,
\end{equation} 
where $:\cdots:$ stands for the normal ordering and $Q_{\sigma,E>\mu_s}=\int^{x_0+v_F \tau}_{x_0} c^\dagger_{\sigma}(x)c_{\sigma}(x) \Theta(E-\mu_S)\, dx$ is nothing else than the probability of presence of a charge $Q$ during a time slot $\tau$ with $E>\mu_S$ and spin projection $\sigma=\uparrow/\downarrow$ along a given direction (\textit{eg} $\hat x$, $\hat z$...). In Appendix B, we discuss in more details the derivation of $Q$ and $\Pi$ depending on the applied filtering, energy or/and spin. In the case of spin filtered detection (for instance spin up with respect to a given direction), this quantity is

\begin{equation}\label{eq:Piup}
\Pi(\tau)=\langle \psi_{\rm out}| : e^{-Q_{\uparrow,E>\mu_s}} :  |\psi_{{\rm out}}\rangle .
\end{equation}
In both cases, the WTD is obtained from the second derivative of the ITP with respect to $\tau$, ${\cal W}(\tau)=\langle \tau \rangle \frac{d^2 \Pi (\tau)}{d\tau^2}$, where $\langle \tau \rangle$ is the mean waiting time given by $1/\langle \tau \rangle=\frac{d\Pi}{d\tau}(\tau=0)$. Eq. (\ref{eq:Pi}) and (\ref{eq:Piup}) are evaluated numerically for both many-body scattering states (\ref{eq:psioutabs}) and (\ref{eq:psioutmbs}) with the same method as Ref. \cite{Haack2014,Albert16}. However, before discussing our results it is useful to recall several established results on WTD in quantum coherent conductors. In Ref. \cite{Albert12,Haack2014}, it was shown that for a single quantum channel with a voltage bias $eV$ (spinless electrons), the scattering quantum state is a train of non interacting fermions whose WTD is approximately the Wigner Surmise

\begin{equation}
  \mathcal W_{WS}(\tau)=\frac{32}{\pi^2}\frac{\tau^2}{\overline \tau ^3}\,\exp\left[-\frac{4}{\pi} \left(\frac{\tau}{\overline \tau}\right)^2\right],
\end{equation}
with $\overline \tau=h/eV$ is the average waiting time, which means that due to Pauli's exclusion principle, electrons are separated in time by $\overline \tau$ on average. An important feature of this WTD is the fact that it vanishes for $\tau\ll \overline \tau$ which is the hallmark of fermionic statistics. If now this stream of electron is partitioned by a scatterer with energy independent transmission coefficient $T$, this WTD is continuously modified until it reaches an exponential form $T \exp(-T \tau/\overline \tau)/\overline \tau$ when $T\ll 1$. This exponential shape is the signature of uncorrelated events since detected electrons are well separated in time and therefore uncorrelated. In this case, the mean waiting time is $\langle\tau\rangle=\overline \tau/T$ and therefore the average current $e / \langle \tau \rangle=\frac{e^2}{h}V T$ in agreement with the so called Landauer's formula \cite{Buttiker2000}. Finally, when spin $1/2$ are considered there are two conducting channel at disposal and the WTD no longer vanishes for small waiting times. At perfect transmission, it is described by the generalized Wigner-Dyson statistics \cite{Haack2014}. 

\section{WTD without spin filtering} 
\begin{figure}
  \includegraphics[width=0.95\linewidth,height=7cm]{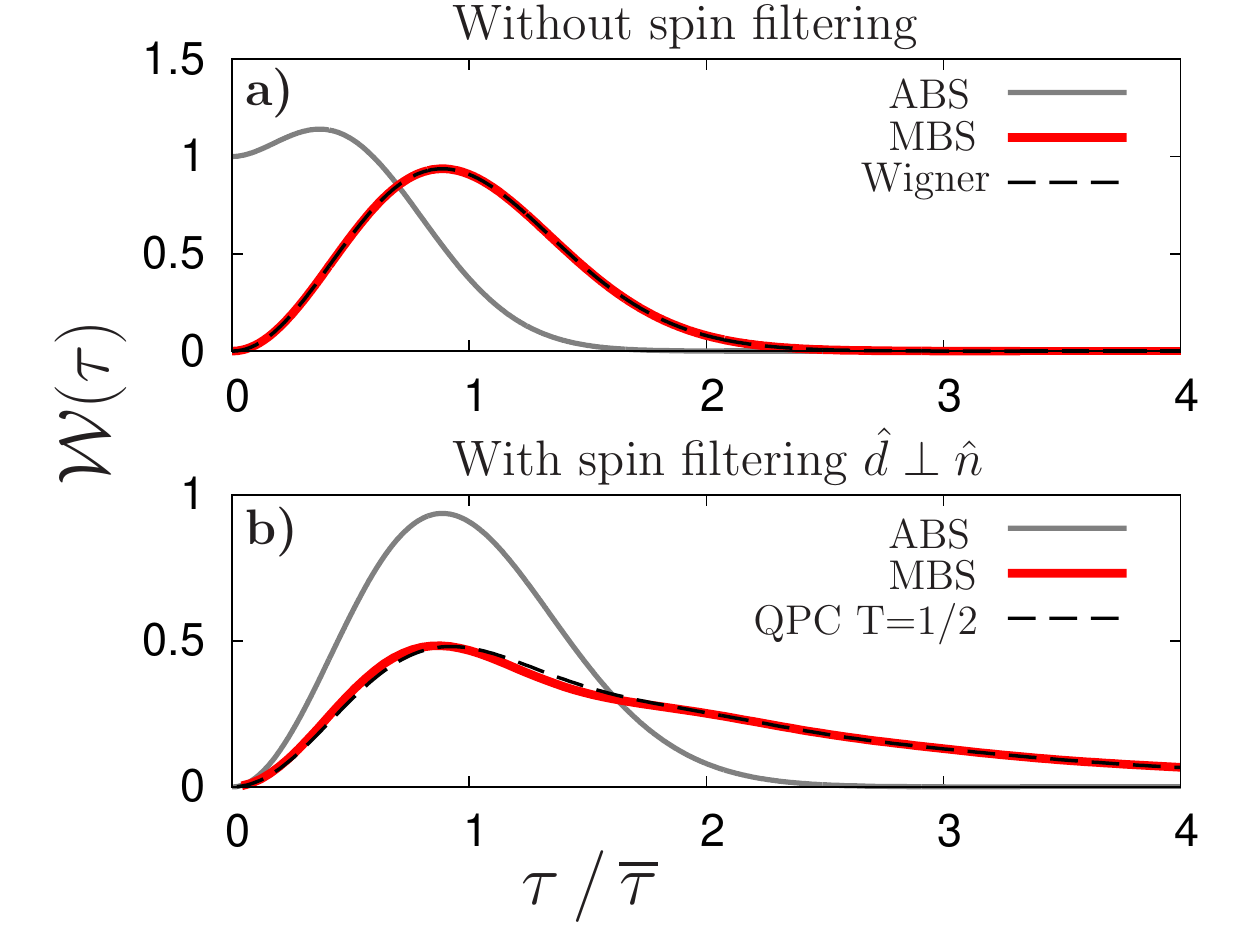}
  \caption{(color online) WTDs versus $\tau/\overline \tau$ of electrons flowing out of the interface without spin filtering {\bf a)} and with spin filtering orthogonal to the majorana polarization {\bf b)}. The solid gray line (resp. red line) corresponds to an NS junction without (resp. with a) MBS (see text) at perfect Andreev reflection. The dashed black line represents the WTD of a single channel normal conductor with transmission probability one ({\bf a)}) and one-half ({\bf b)}) for comparison \cite{Albert12}.}\label{fig_wtd1}
\end{figure}

We start with the simplest situation where the single electron detector is spin insensitive. In the absence of MBS, it was shown \cite{Albert16} that the situation reduces to a stream of one dimensional free electrons with two spin components. Indeed, at perfect Andreev reflection, all the incoming holes are converted into electrons (Andreev mirror) with spin flip and energy between $\mu_S$ and $\mu_S+eV$. The WTD is therefore the one of two perfect and independent quantum channels and is described by the generalized Wigner-Dyson distribution \cite{Haack2014} depicted on Fig. \ref{fig_wtd1}a. The average waiting time is $\langle \tau \rangle=h/2eV$ or in other words the average current is $\frac{2 e^2}{h}V$. On the other hand, in the topological case, the SESAR effect selects only one spin species reducing the possibilities to a single perfect quantum channel (with spin orientation $+\hat n$). As a consequence, the WTD boils down to the so called Wigner surmise \cite{Albert12} also depicted on Fig. \ref{fig_wtd1}a. The average waiting time is $h/eV$ and the average current $\frac{e^2}{h}V$ therefore twice smaller than for the topological case. This is in agreement with the common interpretation that a Majorana behaves as ``half an electron'' \cite{Kitaev2001}.  

We therefore conclude that not only the WTD reproduces well known differences about the average current but also exhibits a qualitative mismatch between the two situations. With MBS, the WTD is exactly zero at $\tau=0$ because of Pauli's exclusion principle whereas it is not in the ABS case since two channels are available \cite{Haack2014}. However, this discrepancy must be visible in any statistical measure of the electronic current like noise, third cumulant and Full Counting Statistics (FCS) in the same way that it is between one and two channel standard mesoscopic conductors.

\section{WTD with spin filtering} We now turn to a richer situation where we assume the single electron detector to be spin sensitive along a direction $\hat d$ and therefore collects only electrons with spin projection $\uparrow_{\hat d}$. The following discussion is basically equivalent to the interpretation of the famous Stern and Gerlach experiment. The key point is that quantum state (\ref{eq:psioutabs}) is spin isotropic whereas (\ref{eq:psioutmbs}) is spin polarized in the $\hat n$ direction. In particular, all possible observables are totally independent of the detector spin orientation in the ABS case. This is in strong contrast with the MBS where, for instance, filtering spin along $\pm \hat n$ leads obviously to orthogonal results. We illustrate this statement on the WTD but it is very important to note that it applies to any other observables such as the spin resolved average current \cite{He2014} or noise \cite{MF_Oreg} and FCS in general.

  For ABS, the single particle detector, whatever its spin orientation $\hat d$, filters one spin species, namely $\uparrow_{\hat d}$. Since they are equally populated and independent, the outgoing state (\ref{eq:psioutabs}) reduces to a single perfect quantum channel and its WTD is therefore Wigner surmise (see Fig. \ref{fig_wtd1}b). This WTD is characterized by a single peak centered around $\tau=\overline \tau$ with broad fluctuations and exactly zero value at zero which is the hallmark of Pauli exclusion principle as already mentioned. When a MBS is present, the precise shape of the WTD crucially depends on the detector spin orientation. Although quite academical, we can start by setting it to $\hat n$. In that case, the detector collects every electron coming from the interface and the WTD is also the one of a single quantum channel. In this situation both ABS and MBS yield the same spin resolved WTD. If we choose now $\hat d=-\hat n$ the detector collects nothing and if $\hat d$ slightly deviates from $-\hat n$ only a few electrons are kept and the WTD is expected to be exponential with rate $\frac{eV}{h} P_{\hat d} $ where $P_{\hat d}$ is the overlap $|\langle \uparrow_{\hat d} | \uparrow_{\hat n}\rangle |^2$.

For arbitrary $\hat d$, the detector will partition the single quantum channel according to spin. The situation is almost formally equivalent to the one of a spinless single quantum channel flowing across a Quantum Point Contact (QPC) with energy independent transmission probability \cite{Albert12}. Here this transmission probability will simply be given by the overlap $P_{\hat d} $ between $|\uparrow_{\hat d}\rangle$ and $|\uparrow_{\hat n}\rangle$.  In the special case where $\hat d\perp \hat n$, the quantum state (\ref{eq:psioutmbs}) is a balanced mixture of $|\uparrow_{\hat d}\rangle$ and $|\downarrow_{\hat d}\rangle$ and then will be filtered exactly like a single quantum channel across a QPC with transmission probability $1/2$. The situation can be implemented experimentally either by tuning $V_z$ just above $\Delta$ and setting $\hat d=\hat z$ or in the limit $V_z\gg\Delta$ where $\hat n=\hat z$ and filtering spin along $x$ or $y$. This is shown on Fig. \ref{fig_wtd1}b where we have evaluated Eq. \ref{eq:Piup} along $\hat z$ right above the topological transition by brute force numerics (limited to a quite small number of basis state (thirteen here) which explains the small discrepancy) and compared it to the expected result with very good agreement.
  
 At this point, it is important to give some explanations on the experimental feasibility. The substrate, namely the heterojunction, has already been fabricated during the quest for Majorana quasi-particle \cite{Mourik,Marcus, Heiblum} and consists of a Rashba nanowire partially in contact with an s-wave superconductor and in presence of a Zeeman field. The crucial point is to detect reflected electrons one by one in the normal part. Although still quite challenging, single electron detection technology is progressing very fast and might become a routine in the near future as reviewed in \cite{Albert2014,Dasenbrook2016}. Otherwise, partial information on waiting times can be extracted from the average current, shot noise or second order coherence function obtained from Hong-Ou-Mandel experiment \cite{Albert2014,Bocquillon2014,Haack2015}. 

\section{Conclusion} We have studied the consequences of the presence or not of a MBS at the interface between a normal and a superconducting conductor on the electronic WTD. When a single electron detector is placed far away from the interface and detects electrons above the superconducting chemical potential $\mu_S$ without spin filtering we observe a clear qualitative distinction between the topological and non topological situations. In addition, we have shown that the non topological situation (ABS) is immune to spin filtering in sharp contrast with the topological one due to SESAR effect. This conclusion is valid for the WTD which makes it a clear fingerprint of MBS but is also true for other quantities like the average current or higher moments of the FCS which can be easier to measure in actual experiments. Extension of this work could be the study of the influence of Coulomb repulsion when the superconducting part is not grounded but floating or the poisoning by another Majorana \cite{Beri2012,Plugge2016} and temperature or disorder effects.\\

\acknowledgments
We are grateful to G. Candela, G. Haack and J. Klinovaja for useful discussions and remarks. The research of D. C. was supported by the Swiss NSF and NCCR QSIT.

\section{Appendix A: Finite width of the outgoing states}

\begin{figure}
  \includegraphics[width=0.95\linewidth]{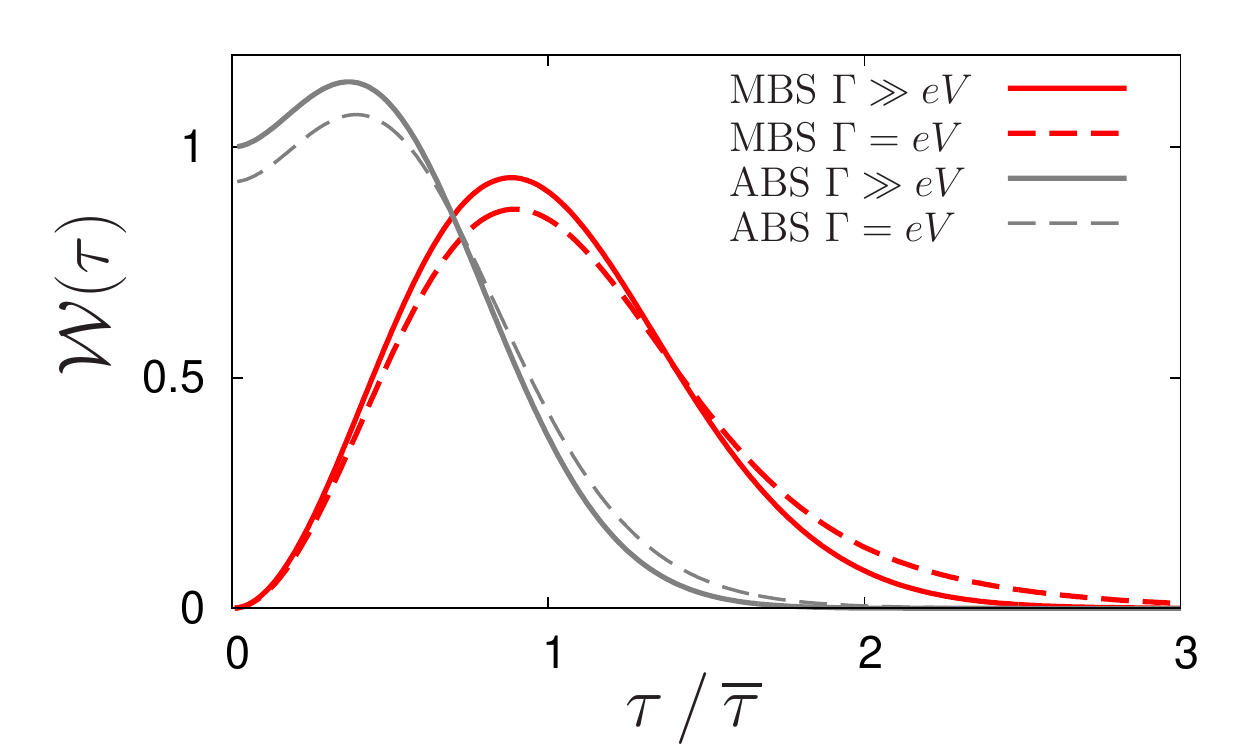}
  \caption{(color online) WTDs versus $\tau/\overline \tau$ of electrons flowing out of the interface of a N/S (gray line) and a N/TS (red line). The dashed and solid line correspond to two different width of the states as mentioned in the legend.}\label{fig_wtd2.pdf}
\end{figure}

We study here the effect of a finite energy width of the Majorana bound state and show that it does not qualitatively change the waiting time distribution. Due to the finite hopping strength at the interface, the Majorana bound state located at zero energy has a finite width $\Gamma$. The main consequence of this is the energy dependence of the Andreev reflection coefficient at the interface which becomes \cite{Ioselevich2013} 
\begin{equation}
R_A(E)=\left| \frac{\Gamma}{E+i\Gamma}\right|^2.
\end{equation}
However, we can recover the same states (2) and (3) as in the main text when taking the limit $eV\ll\Gamma$. Using this assumption, we can calculate the ITP and thus the WTD when the states have a finite width (see Fig. \ref{fig_wtd2.pdf}) using the energy dependent coefficients \cite{Haack2014}.

In fig. \ref{fig_wtd2.pdf}, we can see that the effect of the broadening of the states has no dramatic effects on the WTD. For this reason, we do not focus on this effect in the main text.

\section{Appendix B: Derivation of the Idle Time Probability}

In this appendix we explain how to calculate the idle time probability, following the notations of Ref. \cite{Haack2014}. It is important to note that the energy range of detected particles is assumed to be small enough that the dispersion relation of electrons or holes is linear $E=\hbar v_F k$. In that case, charge measurements over a time window $\Delta t$ are equivalent to charge measurements over a space window $\Delta x/v_F$ thanks to Galilean invariance. The key point to calculate $Q_{\sigma,E}$ is to specify the detection procedure. This includes the possibility to only detect positive/negative energies, spin projection up/down with respect to a given quantization axis. In an actual experiment, the detection can be done by connecting the superconductor or the topological superconductor to two quantum dots instead of a normal metal. By doing so one can filter energy by applying an external gate on the two dots or select the spin by using interacting quantum dots with strong repulsion in order to get rid of the spin degeneracy \cite{Chevallier2011, Recher2001} . Analytically these properties can be implemented easily in the definition of $Q$. This operator can be represented in the basis of the scattering states as

\begin{equation}
  Q_{E>\mu_s}=\int t(k) t^*(k')\frac{e^{i(k-k')v_F\tau}-1}{i(k-k')}\frac{dk}{2\pi}\frac{dk'}{2\pi}
\end{equation}
where the $t(k)$ are the energy dependent transmission amplitudes of a scattering state and may be chosen in that case as $t(k)=1$ if $E(k)>\mu_s$ and $t(k)=0$ if $E(k)<\mu_s$. In order to compute the ITP, the transport window has to be discretized into $N$ energy compartments of size $eV/N$ with corresponding momentum intervals of size $\kappa=\frac{eV}{N\hbar v_F}$ with $v_F=\frac{\hbar k_F}{m}$ is the Fermi velocity defined with $m$ the electron mass. Using this discretization leads to the following matrix elements for $Q_{E>\mu_s}$ in the large $N$ limit

\begin{equation}
[Q]_{m,n}= \frac{ \kappa t^*_{\kappa m} t_{\kappa n}}{\pi}e^{-\frac{i}{2}\kappa( n- m) v_F\tau} \frac{\textrm{sin}\left[\frac{(\kappa n-\kappa m) v_F \tau}{2}\right]}{(\kappa n-\kappa m)}
\end{equation}

with $m,n=1,....,N$. From this definition of the ITP and when the average is taken over a Slater determinant of free fermions, the WTD can be cast as a determinant of the form \cite{Haack2014,Dasenbrook}

\begin{equation}
\Pi (\tau)=\textrm{det}(1-Q_{\tau}),
\end{equation}
which can be evaluated with a computer. Then, it is straigthforward to extend this detection procedure to a spin selective one by setting spin dependent transmission coefficients.


\begin{thebibliography}{0}

\bibitem{Fu} 
  \Name{Fu L. \and Kane C. L.}
  \REVIEW{Phys. Rev. Lett.}{100}{2008}{096407}.

\bibitem{MF_Sato} 
  \Name{Sato M. \and Fujimoto S.}
  \REVIEW{Phys. Rev. B}{79}{2009}{094504}.

\bibitem{MF_Sarma} 
  \Name{Lutchyn R.M., Sau J. D. \and Das Sarma S.}
  \REVIEW{Phys. Rev. Lett.}{105}{2010}{077001}.

\bibitem{MF_Oreg} 
  \Name{Oreg Y.,  Refael G. \and  v. Oppen F.}
  \REVIEW{Phys. Rev. Lett.}{105}{2010}{177002}.

\bibitem{alicea_majoranas_2010} 
  \Name{Alicea J.}
  \REVIEW{Phys. Rev. B}{81}{2010}{125318}.

\bibitem{potter_majoranas_2011} 
  \Name{Potter A. C. \and Lee P. A.}
  \REVIEW{Phys. Rev. B}{83}{2011}{094525}.

\bibitem{Klinovaja_CNT} 
  \Name{Klinovaja J., Gangadharaiah S. \and Loss D.}
  \REVIEW{Phys. Rev. Lett.}{108}{2012}{196804}.

\bibitem{Pascal} 
  \Name{Chevallier D., Sticlet D., Simon P. \and Bena C.}
  \REVIEW{Phys. Rev. B}{85}{2012}{235307}.

\bibitem{Bena_MF} 
  \Name{Sticlet D., Bena C., \and Simon P.}
  \REVIEW{Phys. Rev. Lett.}{108}{2012}{096802}.

\bibitem{Rotating_field} 
  \Name{Klinovaja J., Stano P. \and Loss D.}
  \REVIEW{Phys. Rev. Lett}{109}{2012}{236801}.

\bibitem{Ali} 
  \Name{Nadj-Perge S.,Drozdov I. K.,Bernevig B. A., \and Yazdani A.}
  \REVIEW{Phys. Rev. B}{88}{2013}{020407(R)}.

\bibitem{RKKY_Basel} 
  \Name{Klinovaja J., Stano P., Yazdani A., \and Loss D.}
  \REVIEW{Phys. Rev. Lett.}{111}{2013}{186805}.

\bibitem{RKKY_Simon} 
  \Name{Braunecker B. \and Simon P.}
  \REVIEW{Phys. Rev. Lett.}{111}{2013}{147202}.

\bibitem{RKKY_Franz} 
  \Name{Vazifeh M. M. \and Franz M.}
  \REVIEW{Phys. Rev. Lett.}{111}{2013}{206802}.

\bibitem{Pientka} 
  \Name{Pientka F., Glazman L. I. \and v. Oppen F.}
  \REVIEW{Phys. Rev. B}{88}{2013}{155420}.

\bibitem{Ojanen} 
  \Name{Poyh\"onen K., Weststr\"om A., R\"ontynen J. \and Ojanen T.}
  \REVIEW{Phys. Rev. B}{89}{2004}{115106}.

\bibitem{Fra} 
  \Name{Maier F., Klinovaja J. \and  Loss D.}
  \REVIEW{Phys. Rev. B}{90}{2014}{195421}.

\bibitem{Carlos} 
  \Name{Vernek E., Penteado P. H., Seridonio A. C. \and Egues J. C.}
  \REVIEW{Phys. Rev. B}{89}{2014}{165314}.

\bibitem{Alicea} 
  \Name{Alicea J.}
  \REVIEW{Rep. Prog. Phys.}{75}{2012}{076501}.

\bibitem{Beenakker} 
  \Name{Beenakker C. W. J.}
  \REVIEW{Annu. Rev. Con. Mat. Phys.}{4}{2013}{113}.


\bibitem{Mourik} 
  \Name{Mourik V., Zuo K., Frolov S. M., Plissard S. R., Bakkers E. P. A. M. \and Kouwenhoven L. P.}
  \REVIEW{Science}{336}{2012}{1003}.

\bibitem{Marcus} 
  \Name{Churchill H. O. H., Fatemi V., Grove-Rasmussen K., Deng M. T., Caroff P., Xu H. Q. \and Marcus C. M.}
  \REVIEW{Phys. Rev. B}{87}{2013}{241401}.

\bibitem{Heiblum} 
  \Name{Das A., Ronen Y., Most Y., Oreg Y., Heiblum M. \and Shtrikman H.}
  \REVIEW{Nat. Phys.}{8}{2012}{887}.

\bibitem{Us} 
  \Name{Rokhinson L. P., Liu X. \and Furdyna J. K. }
  \REVIEW{Nat. Phys.}{8}{2012}{795}.

\bibitem{DeFranceschi} 
  \Name{Lee E. J. H., Jiang X., Aguado R., Katsaros G., Lieber C. M. \and De Franceschi S.}
  \REVIEW{Phys. Rev. Lett.}{109}{2012}{186802}.

\bibitem{Liu2012} 
  \Name{Liu J., Potter A. C., Law K. T. \and Lee P. A.}
  \REVIEW{Phys. Rev. Lett.}{109}{2012}{267002}.

\bibitem{Atland2012} 
  \Name{Bagrets D. \and Altland A.}
  \REVIEW{Phys. Rev. Lett.}{109}{2012}{227005}.

\bibitem{Pientka2012} 
  \Name{Pientka F., Kells G., Romito A., Brouwer P. W. \and von Oppen F.}
  \REVIEW{Phys. Rev. Lett.}{109}{2012}{227006}.

\bibitem{Pikulin2012} 
  \Name{Pikulin D. I., Dahlhaus J. P., Wimmer M., Schomerus H. \and Beenakker C. W. J.}
  \REVIEW{New. J. Phys.}{14}{2012}{125011}.

\bibitem{Klinovaja} 
  \Name{Klinovaja J. \and Loss D.}
  \REVIEW{Phys. Rev. B}{86}{2012}{085408}.

\bibitem{Ramon} 
  \Name{Prada E., San-Jose P. \and Aguado R.}
  \REVIEW{Phys. Rev. B}{86}{2012}{180503(R)}.

\bibitem{Fu2009}
\Name{Fu L. \and Kane C.}
  \REVIEW{Phys. Rev. B}{79}{2009}{161408(R)}.


\bibitem{Rubbert} 
  \Name{Rubbert S., Akhmerov A. R.}
  \REVIEW{Phys. Rev. B}{94}{2016}{115430}.

\bibitem{Beenakker2016} 
  \Name{Beenakker C. W. J. \and Kowenhoven L.}
  \REVIEW{Nat. Phys.}{12}{2016}{618}.


\bibitem{He2014} 
  \Name{He J. J., Ng T. K., Lee P. A. \and Law K. T.}
  \REVIEW{Phys. Rev. Lett.}{112}{2014}{037001}.

\bibitem{Haim2015}
  \Name{Haim A., Berg E., von Oppen F. \and Oreg Y.}
  \REVIEW{Phys. Rev. Lett.}{114}{2015}{166406}.

\bibitem{Brandes2008} 
  \Name{Brandes T.}
  \REVIEW{Ann. Phys. (Berlin)}{17}{2008}{477}.

\bibitem{Albert11} 
  \Name{Albert M., C. Flindt C. \and M. B\"{u}ttiker M.}
  \REVIEW{Phys. Rev. Lett.}{107}{2011}{086805}.

\bibitem{Albert12} 
  \Name{Albert M., Haack G., Flindt C. \and B\"uttiker M.}
  \REVIEW{Phys. Rev. Lett.}{108}{2012}{186806}.

\bibitem{FlindtThomas} 
  \Name{Thomas K. H. \and C. Flindt C.}
  \REVIEW{Phys. Rev. B}{87}{2013}{121405(R)}.

\bibitem{Dasenbrook} 
  \Name{Dasenbrook D., Flindt C., \and M. B\"uttiker M.}
  \REVIEW{Phys. Rev. Lett.}{112}{2014}{146801}.

\bibitem{Rajabi2013} 
  \Name{Rajabi L., P\"oltl C. \and Governale M.}
  \REVIEW{Phys. Rev. Lett.}{111}{2013}{067002}.

\bibitem{FlindtThomas2} 
  \Name{Thomas K. H. \and Flindt C.}
  \REVIEW{Phys. Rev. B}{89}{2014}{245420}.

\bibitem{Sothmann} 
  \Name{Sothmann B.}
  \REVIEW{Phys. Rev. B}{90}{2014}{155315}.

\bibitem{Albert2014} 
  \Name{Albert M. \and Devillard P.}
  \REVIEW{Phys. Rev. B}{90}{2014}{035431}.

\bibitem{Haack2014} 
  \Name{Haack G., Albert M. \and C. Flindt C.}
  \REVIEW{Phys. Rev. B}{90}{2014}{205429}.

\bibitem{Hofer16} 
  \Name{Hofer P. P., Dasenbrook D. \and C. Flindt C.}
  \REVIEW{Physica E (2015), doi: 10.1016/j.physe.2015.08.034}{}{}{}.

\bibitem{Albert16} 
  \Name{Albert M., Chevallier D. \and Devillard P.}
  \REVIEW{Physica E}{76}{2016}{209}.

\bibitem{Dasenbrook2016} 
  \Name{Dasenbrook D.\and C. Flindt C.}
  \REVIEW{Phys. Rev. B}{93}{2016}{245409}.

\bibitem{Brandes2016} 
  \Name{Brandes T. and Emary C.}
  \REVIEW{ Phys. Rev. E}{93}{2016}{042103}.

\bibitem{PikulinJETP}
   \Name{Pikulin D. I.\and Nazarov Y.}
   \REVIEW{JETP Lett.}{94}{2012}{693Ð697}.

\bibitem{PikulinPRB}
   \Name{Pikulin D. I. \and Nazarov Y.}
   \REVIEW{Phys. Rev. B}{87}{2013}{235421}.

\bibitem{Jose2016}
   \Name{San Jose P., Cayao J., Prada E. \and Aguado R.}
   \REVIEW{Sci. Rep.}{6}{2016}{21427}.

\bibitem{Buttiker2000} 
  \Name{Blanter Ya. M. \and B\"{u}ttiker M.}
  \REVIEW{Phys. Rep.}{336}{2000}{1}.

\bibitem{Kitaev2001} 
  \Name{Kitaev A. Y.}
  \REVIEW{Phys. Usp.}{44}{2001}{131}.

\bibitem{Bocquillon2014} 
  \Name{E. Bocquillon {\it et al}}
  \REVIEW{Annalen der Physik}{526}{2014}{1}.

\bibitem{Haack2015} 
  \Name{Haack G., Steffens A., J. Eisert J. \and H\"ubener R.}
  \REVIEW{New J. Phys.}{17}{2015}{113024}.

\bibitem{Beri2012} 
  \Name{Beri B. \and Cooper N. R.}
  \REVIEW{Phys. Rev. Lett.}{109}{2012}{156803}.

\bibitem{Plugge2016} 
  \Name{Plugge S., Zazunov A., Erikson E., Tsvelik A. M. \and Egger R.}
  \REVIEW{Phys. Rev. B}{93}{2016}{104524}.
  
 \bibitem{Ioselevich2013}
  \Name{Ioselevich. P,  \and Feigel'man M. V.}
  \REVIEW{New. J. Phys.}{15}{2013}{055011}.
  
 \bibitem{Chevallier2011}
  \Name{Chevallier. D, Rech J., Jonckheere T. \and Martin T.}
  \REVIEW{Phys. Rev. B}{83}{2011}{125421}.
  
 \bibitem{Recher2001}
  \Name{Recher. P, Sukhorukov E. V. \and Loss D.}
  \REVIEW{Phys. Rev. B}{63}{2001}{165314}.


\end{thebibliography}
\end{document}